\documentclass[sn-mathphys]{sn-jnl}% Math and Physical Sciences Reference Style

\usepackage[T1]{fontenc}
\usepackage[separate-uncertainty=true,multi-part-units=single,per-mode=repeated-symbol]{siunitx}
\usepackage{graphicx}
\usepackage{gensymb}
\usepackage{xspace}
\usepackage{slashbox}

\jyear{2024}
\newcommand{\Hbar}{\overline{H}\xspace}
\newcommand{\pbar}{\overline{p}\xspace}
\newcommand{\aegis}{AE$\bar{\text{g}}$IS\xspace}

\begin{document}

% \title[Real-time antiproton annihilation detection with unprecedented resolution to enable high precision gravitational measurements]{Real-time antiproton annihilation detection with unprecedented resolution to enable high precision gravitational measurements}

\title[Real-time antiproton annihilation vertexing with sub-micron resolution]{Real-time antiproton annihilation vertexing with sub-micron resolution}

%%=============================================================%%
%% Prefix	    -> \pfx{Dr}
%% Particle	    -> \spfx{van der} -> surname prefix
%% Suffix	    -> \sfx{IV}
%% Degrees	    -> \dgr{MSc, PhD}

% \input{Authors2}

\author[1]{\fnm{M.} \sur{Berghold}}\email{michael.berghold@frm2.tum.de}
\equalcont{These authors contributed equally to this work.}

\author[2]{\fnm{D.} \sur{Orsucci}}\email{davide.orsucci@dlr.de}

\author*[1,3,4]{\fnm{F.} \sur{Guatieri}}\email{francesco.guatieri@frm2.tum.de}
\equalcont{These authors contributed equally to this work.}

\author[5]{S. Alfaro}
\author[6]{M. Auzins}
\author[7]{B. Bergmann}
\author[7]{P. Burian}
\author[3,4]{R. S. Brusa}
\author[8]{A. Camper}
\author[3,4]{R. Caravita}
\author[9,10]{F. Castelli}
\author[11]{G. Cerchiari}
\author[12]{R. Ciury{\l{}}o}
\author[3,4]{A. Chehaimi}
\author[9,13]{G. Consolati}
\author[14]{M. Doser}
\author[15]{K. Eliaszuk}
\author[3,4]{R. Ferguson}
\author[14]{M. Germann}
\author[15]{A. Giszczak}
\author[14]{L. T. Gl\"{o}ggler}
\author[15]{{\L.} Graczykowski}
\author[14]{M. Grosbart}
%\author[3,4]{F. Guatieri}
\author[14,16]{N. Gusakova}
\author[14]{F. Gustafsson}
\author[14]{S. Haider}
\author[14,17]{S. Huck}
\author[1]{C. Hugenschmidt}
\author[15]{M. A. Janik}
\author[15]{T. Januszek}
\author[18]{G. Kasprowicz}
\author[15]{K. Kempny}
\author[19]{G. Khatri}
\author[12]{\L. K\l osowski}
\author[15]{G. Kornakov}
\author[6,14]{V. Krumins}
\author[15]{L. Lappo}
\author[12]{A. Linek}
\author[3,4]{S. Mariazzi}
\author[20,21]{P. Moskal}
\author[15]{D. Nowicka}
\author[20,21]{P. Pandey}
\author[15,22]{D. P{\k e}cak}
\author[3,4]{L. Penasa}
\author[23]{V. Petracek}
\author[12]{M. Piwi\'nski}
\author[7]{S. Pospisil}
\author[3,4]{L. Povolo}
\author[9]{F. Prelz}
\author[24]{S. A. Rangwala}
\author[13,14,25]{T. Rauschendorfer}
\author[26,27]{B. S. Rawat}
\author[26]{B. Rien\"{a}cker}
\author[26]{V. Rodin}
\author[8]{O. M. R{\o}hne}
\author[8]{H. Sandaker}
\author[20,21]{S. Sharma}
\author[7]{P. Smolyanskiy}
\author[22]{T. Sowi\'nski}
\author[15]{D. Tefelski}
\author[14]{T. Vafeiadis}
\author[3,4,13,14]{M. Volponi}
\author[26,27]{C. P. Welsch}
\author[12]{M. Zawada}
\author[15]{J. Zielinski}
\author[28,29]{N. Zurlo}

\affil[1]{\orgdiv{Research Neutron Source Heinz Maier-Leibnitz (FRM II)}, \orgname{Technical University of Munich}, \orgaddress{\street{Lichtenbergstr. 1}, \city{Garching bei M\"unchen}, \postcode{85748}, \state{Bayern}, \country{Germany}}}
\affil[2]{\orgdiv{Institute of Communications and Navigation}, \orgname{German Aerospace Centre (DLR)}, \orgaddress{Münchener Str. 20, 82234 We\ss ling, Germany}}
\affil[3]{Department of Physics, University of Trento, via Sommarive 14, 38123 Povo, Trento, Italy}
\affil[4]{TIFPA/INFN Trento, via Sommarive 14, 38123 Povo, Trento, Italy}
\affil[5]{University of Siegen, Department of Physics, Walter-Flex-Straße 3, 57072 Siegen, Germany}
\affil[6]{University of Latvia, Department of Physics Raina boulevard 19, LV-1586, Riga, Latvia}
\affil[7]{Institute of Experimental and Applied Physics, Czech Technical University in Prague, Husova 240/5, 110 00, Prague 1, Czech Republic}
\affil[8]{Department of Physics, University of Oslo, Sem Sælandsvei 24, 0371 Oslo, Norway}
\affil[9]{INFN Milano, via Celoria 16, 20133 Milano, Italy}
\affil[10]{Department of Physics ``Aldo Pontremoli'', University of Milano, via Celoria 16, 20133 Milano, Italy}
\affil[11]{Institut f\"{u}r Experimentalphysik, University of Innsbruck, Technikerstrasse 25, 6020 Innsbruck, Austria}
\affil[12]{Institute of Physics, Faculty of Physics, Astronomy, and Informatics, Nicolaus Copernicus University in Toru\'n, Grudziadzka 5, 87-100 Toru\'n, Poland}
\affil[13]{Department of Aerospace Science and Technology, Politecnico di Milano, via La Masa 34, 20156 Milano, Italy}
\affil[14]{Physics Department, CERN, 1211 Geneva 23, Switzerland}
\affil[15]{Warsaw University of Technology, Faculty of Physics, ul. Koszykowa 75, 00-662, Warsaw, Poland}
\affil[16]{Department of Physics, NTNU, Norwegian University of Science and Technology, Trondheim, Norway}
\affil[17]{Institute for Experimental Physics, Universit\"{a}t Hamburg, 22607, Hamburg, Germany}
\affil[18]{Warsaw University of Technology, Faculty of Electronics and Information Technology, ul. Nowowiejska 15/19, 00-665 Warsaw, Poland}
\affil[19]{Systems Department, CERN, 1211 Geneva 23, Switzerland}
\affil[20]{Marian Smoluchowski Institute of Physics, Jagiellonian University, Krak\'ow, Poland}
\affil[21]{Centre for Theranostics, Jagiellonian University, Krak\'ow, Poland}
\affil[22]{Institute of Physics, Polish Academy of Sciences, Aleja Lotnikow 32/46, PL-02668 Warsaw, Poland}
\affil[23]{Czech Technical University, Prague, Brehov\'a 7, 11519 Prague 1, Czech Republic}
\affil[24]{Raman Research Institute, C. V. Raman Avenue, Sadashivanagar, Bangalore 560080, India}
\affil[25]{Felix Bloch Institute for Solid State Physics, Universit\"{a}t Leipzig, 04103 Leipzig, Germany}
\affil[26]{Department of Physics, University of Liverpool, Liverpool L69 3BX, UK}
\affil[27]{The Cockcroft Institute, Daresbury, Warrington WA4 4AD, UK}
\affil[28]{INFN Pavia, via Bassi 6, 27100 Pavia, Italy}
\affil[29]{Department of Civil, Environmental, Architectural Engineering and Mathematics, University of Brescia, via Branze 43, 25123 Brescia, Italy}

\abstract{
The primary goal of the \aegis experiment is to precisely measure the free fall of antihydrogen within Earth's gravitational field. To this end, a cold ($\approx \SI{50}{\kelvin}$) antihydrogen beam has to pass through two grids forming a moir\'e deflectometer before annihilating onto a position-sensitive detector, which shall determine the vertical position of the annihilation vertex relative to the grids with micrometric accuracy. Here we introduce a vertexing detector based on a modified mobile camera sensor and experimentally demonstrate that it can measure the position of antiproton annihilations with an accuracy of $0.62^{+0.40}_{-0.22}\,$\SI{}{\micro\meter}, which represents a 35-fold improvement over the previous state-of-the-art for real-time antiproton vertexing. Importantly, these antiproton detection methods are directly applicable to antihydrogen. Moreover, the sensitivity to light of the sensor enables the in-situ calibration of the moir\'e deflectometer, significantly reducing systematic errors. This sensor emerges as a breakthrough technology for achieving the \aegis scientific goals and has been selected as the basis for the development of a large-area detector for conducting antihydrogen gravity measurements.
}



\maketitle

\section{Introduction}\label{introduction}

General relativity currently stands as the prevailing paradigm for understanding gravity and any departure from it would require a fundamental reevaluation of our understanding of physics. At its core lies the Weak Equivalence Principle (WEP), postulating the equality of gravitational and inertial mass and thus the universality of free fall. The WEP has been tested across various materials to very high precision through Eötvös-type experiments~\cite{wagner2012torsion,MICROSCOPE}. Testing the WEP with antimatter presents a formidable challenge due to the difficulty of producing abundant low-energy antiparticles and the susceptibility to stray electromangnetic fields~\cite{witteborn1968experiments,HOLZSCHEITER1993}. The idea that antimatter should adhere to the WEP is supported both by theoretical considerations, such as conservation of energy combined with the observed blue-shift of light in gravitational fields~\cite{morrison1958approximate}; and by indirect experimental evidence, such as the lack of anomalous regeneration of $K$-short kaons~\cite{Good1961}, the precise measurement of the antiproton cyclotron frequency~\cite{borchert2022}, and the times of arrival of neutrinos and antineutrinos from the supernova SN1987a~\cite{LoSecco}. However, such arguments are not conclusive~\cite{nieto1991arguments} and a direct experimental confirmation was lacking until recently.

% \note{We are probably exceeding the word count limit, so we could take out the first paragraph.}

Over the past decade, several experimental collaborations have been established with the goal of measuring the acceleration of antimatter in Earth's gravitational field, with most of these efforts taking place at the antimatter factory at CERN. These have culminated in an experiment by the ALPHA-g collaboration~\cite{Alpha_g}, which has measured the gravitational acceleration of antihydrogen ($\Hbar$) to be 
$g_{\Hbar} = (0.75 \pm 0.13_{\text{stat} + \text{sys}} \pm 0.16_\text{sim}) \, g$
where $g$ is the local gravitational acceleration on ordinary matter. Zero gravitational pull ($g_{\Hbar} = 0$) is disfavored compared to ordinary gravity ($g_{\Hbar} = g$) with a Bayes factor of $2.9 \cdot 10^{-4}$, while anti-gravity ($g_{\Hbar} = -g$) is conclusively ruled out.

The next steps in advancing the field involve replicating the measurement and then increasing the accuracy to subject the WEP to more stringent tests. Theoretical models that presuppose that antimatter and energy respond differently to gravitational fields~\cite{bondi1957negative, schiff1959gravitational} result in potential deviations from the WEP at the percent level for hadronic matter, including antiprotons and antihydrogen. In fact, according to the Standard Model most of the mass of an antiproton, around $\SI{938}{\mega\electronvolt \per c\squared}$, stems from its internal binding energy~\cite{wilczek2012origins}, whereas the masses of the three constituent antiquarks add up to around $\SI{9}{\mega\electronvolt \per c\squared}$~\cite{pdg}. Since it is established that binding energy responds positively to gravity~\cite{wagner2012torsion,MICROSCOPE}, the deviation $\Delta g = g_{\Hbar} - g$ would not be expected to exceed 1\%~\cite{fischler2008direct, alves2009experimental}.

Achieving a 1\% precision on $g_{\Hbar}$ is a challenging but realistic goal pursued by the competing experiments ALPHA-g~\cite{alphag1}, GBAR~\cite{gbar1} and \aegis~\cite{aegis1} at CERN. These employ substantially different methods to measure the effect of gravity on antihydrogen, each one requiring the development of bespoke technologies and the solution of distinctive technological challenges. The gravitational measurement at \aegis is based on a moir\'e deflectometer (already tested with antiprotons~\cite{aghion2014moire}) in which the particles pass through two material grids before annihilating onto a position-sensitive detector. The vertical deflection of the moir\'e pattern, given by $\delta y = -t^2 g_{\Hbar}$ for a time of flight $t$ between the grids, is expected to be tens of micrometers for the cold antihydrogen beam of \aegis. Therefore, a crucial requirement for high-precision gravity measurements at \aegis is the real-time detection of the position of $\Hbar$ annihilations with micrometric accuracy. Previously considered detection technologies included nuclear photographic emulsions~\cite{kimura2013development, aghion2013prospects, pistillo2015emulsion} and bespoke silicon detectors~\cite{aghion2014detection}, such as Timepix3~\cite{pacifico2016direct, holmestad2018data}. Nuclear emulsions feature a high resolution, approaching $\SI{1}{\micro\meter}$~\cite{kimura2013development}, but, due to the lack of real-time detection, it is challenging to perform on-line monitoring of the position of the photographic plate relative to the grids. On the other hand, real-time detection is possible with silicon detectors, but the ones so far available had a significantly lower resolution. For instance, the pixels of Timepix3 and of Timepix4 are $\qtyproduct{55x55}{\micro\meter}$ in size~\cite{poikela2014timepix3, llopart2022timepix4}.

In this work we introduce a groundbreaking $\Hbar$ detection technology which, for the first time, meets all the requirements to reach a 1\% precision measurement of $g_{\Hbar}$ at \aegis. This is based on the Sony IMX219, a commercial CMOS optical image sensor with 8 megapixels and a $\qtyproduct{3.67x2.76}{\milli\meter}$ sensitive area~\cite{Datasheet, ReverseEngineering}, originally developed for mobile applications and which has already been shown to be capable of imaging low-energy positrons with exceptional efficiency and resolution~\cite{Berghold2023}. Its $\qtyproduct{1.12x1.12}{\micro\meter}$ pixels\footnote{In this paper, we use the convention $\SI{1}{px}=\SI{1.12}{\micro\meter}.$} are fifty times smaller than that of Timepix3 and of size similar to that of nuclear emulsions grains. We have now experimentally observed that the annihilation of antiprotons ($\pbar$) on the surface of the CMOS sensor results in the secondary charged particles leaving detectable signals in the exposed images. By reconstructing the vertex lying at the intersection of the tracks, the position of the antiproton annihilation can be determined with sub-micron accuracy. Since antihydrogen is an antiproton having only an extra positron bound to it, our detector is expected to be an antihydrogen annihilation vertexer as well and thus enable $\Hbar$ gravity measurements.

\section{Results}\label{results}

\subsection{Detection}

\begin{figure*}[t]
    \center
    \includegraphics[width=0.99\textwidth]{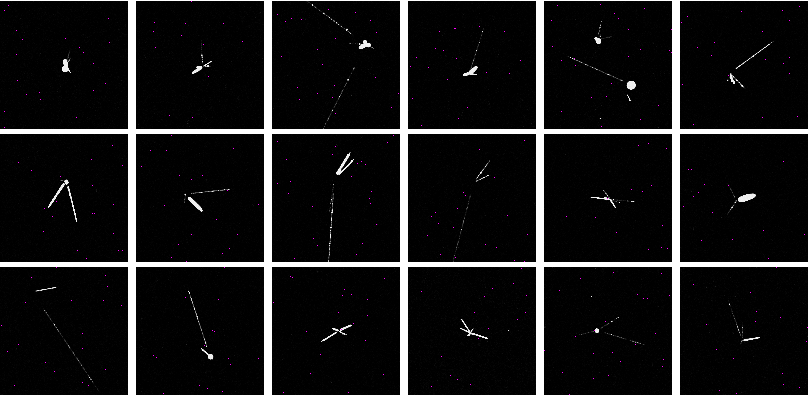}
    \caption{Curated selection of antiproton annihilation events as imaged by the CMOS sensor. The observed shapes are similar to those recorded by the Timepix3 detector~\cite{pacifico2016direct, holmestad2018data}, albeit at a scale approximately fifty times smaller. Pixels marked in magenta have been deemed non-functional by taking background images.}
    \label{fig:StarShape}
\end{figure*}

Low-energy antiprotons impinging on matter are trapped, through the Day-Snow-Sucher mechanism~\cite{DaySnowSucher}, by atomic nuclei and then quickly annihilate with either a proton or a neutron, releasing almost $\SI{1.9}{\giga\electronvolt}$ of energy. Through the interaction of the constituents quarks $p\pbar$ and $n\pbar$ annihilations result in the emission of 2 to 8 pions ($\pi^-,\pi^0,\pi^+$) and other particles~\cite{von1987interaction, richard2020antiproton}. The annihilation in heavy nuclei can result in the recoil of some of the adjacent protons and neutrons and each of these, if the trajectory crosses the nucleus, can also pick-up other nucleons and result in the emission of deuterons, tritium, helium and other heavier nuclear fragments~\cite{clover1982low, von1987interaction}. The secondary particle branching ratios and spectra had been experimentally characterised at the Low Energy Anti-Proton Ring facility between 1982 and 1996~\cite{walcher1988experiments, hofmann1990charged, amsler1991low, polster1995light, amsler1998proton, amsler2019nucleon}.

We have detected $\pbar$ annihilation events by repeatedly exposing the CMOS sensor to a low-intensity beam of antiprotons. Over the course of 8 days a total of 1576 images were exposed to collect antiproton events. Each shot was preceded by an exposure without antiprotons for background subtraction. Within the recorded images a total of 2601 $\pbar$ annihilation events were manually identified. Annihilations appear as star-shaped events with multiple prongs emanating from one single primary vertex (see Figure~\ref{fig:StarShape}). 

The sensor was prepared by removal of the microlenses and Bayer filter using the procedure described in~\cite{Berghold2023}, but leaving the passivation and wire grid layers in place. At the employed implantation energy of \SI{8}{\kilo eV} most of the antiprotons annihilate within $\SI{0.25}{\micro\meter}$ from the surface, directly above the sensor's sensitive volume, which is thus intersected by half of the emitted secondary particles. Some of these are energetic charged particles that generate electron-hole pairs in the silicon, a process that results in a signal similar to the one produced by visible light which the detector integrated electronics is designed to pick up~\cite{Datasheet, Berghold2023, ReverseEngineering}. Neutrons, neutral pions, photons and Auger electrons are also featured prominently among the annihilation products, but are not expected to produce a significant signal. Tertiary particles, e.g.\ due to in-flight decay of a secondary, are rarely observed.

\subsection{Prong identification and detection efficiency}

\begin{table}
    \centering
    \begin{tabular}{|c|c|rrrrrrr|}
        \hline
        $\boldsymbol{\alpha}$ &
        \backslashbox{$\boldsymbol{p}$}{$\boldsymbol{\pi}$} & \textbf{0} & \textbf{1} & \textbf{2} & \textbf{3} & \textbf{4} & \textbf{5} & \textbf{6}  \\ \hline
            \multirow{7}{*}{\textbf{0}}
            & \textbf{0} & --- & --- & 71  & 34  & 9   & 1   & 0   \\
            & \textbf{1} & --- & 116 & 149 & 61  & 24  & 5   & 2   \\
            & \textbf{2} & 81  & 107 & 108 & 64  & 17  & 5   & 0   \\
            & \textbf{3} & 33  & 51  & 59  & 31  & 11  & 5   & 1   \\
            & \textbf{4} & 4   & 16  & 12  & 9   & 2   & 3   & 0   \\
            & \textbf{5} & 0   & 0   & 1   & 1   & 1   & 1   & 0   \\
            & \textbf{6} & 0   & 0   & 0   & 0   & 1   & 0   & 0   \\ \hline 
        
            \multirow{6}{*}{\textbf{1}} 
            & \textbf{0} & ---  & 101 & 44  & 17  & 5   & 0   & 0   \\
            & \textbf{1} & 289 & 247 & 88  & 23  & 5   & 0   & 0   \\
            & \textbf{2} & 156 & 153 & 67  & 13  & 6   & 0   & 0   \\
            & \textbf{3} & 29  & 43  & 30  & 11  & 5   & 1   & 0   \\
            & \textbf{4} & 6   & 4   & 3   & 1   & 3   & 0   & 0   \\
            & \textbf{5} & 0   & 0   & 1   & 0   & 0   & 0   & 0   \\ \hline
         
            \multirow{4}{*}{\textbf{2}} 
            & \textbf{0} & 21  & 14  & 5   & 1   & 0   & 0   & 0   \\
            & \textbf{1} & 23  & 30  & 6   & 6   & 2   & 0   & 0   \\
            & \textbf{2} & 17  & 10  & 5   & 1   & 0   & 0   & 1   \\
            & \textbf{3} & 1   & 2   & 2   & 0   & 0   & 0   & 0   \\ \hline
         
            \multirow{3}{*}{\textbf{3}} 
            & \textbf{0} & 5   & 0   & 0   & 0   & 0   & 0   & 0   \\
            & \textbf{1} & 0   & 1   & 0   & 0   & 0   & 0   & 0   \\
            & \textbf{2} & 0   & 0   & 1   & 0   & 0   & 0   & 0   \\ \hline
        \end{tabular} \vspace{2mm}
    \caption{Number of recorded events sorted by prong composition. Prongs appearing as \textit{ellipses} are identified as alpha particles or other multi-nucleon fragments ($\boldsymbol{\alpha}$), \textit{thick tracks} are identified as protons ($\boldsymbol{p}$) and \textit{thin tracks} are identified as charged pions ($\boldsymbol{\pi}$). Annihilation events featuring zero or one prongs cannot be traced and are excluded from the analysis (---).}
    \label{tab:Composition}
\end{table}

\begin{figure*}[t]
    \center
    \includegraphics[width=0.60\textwidth]{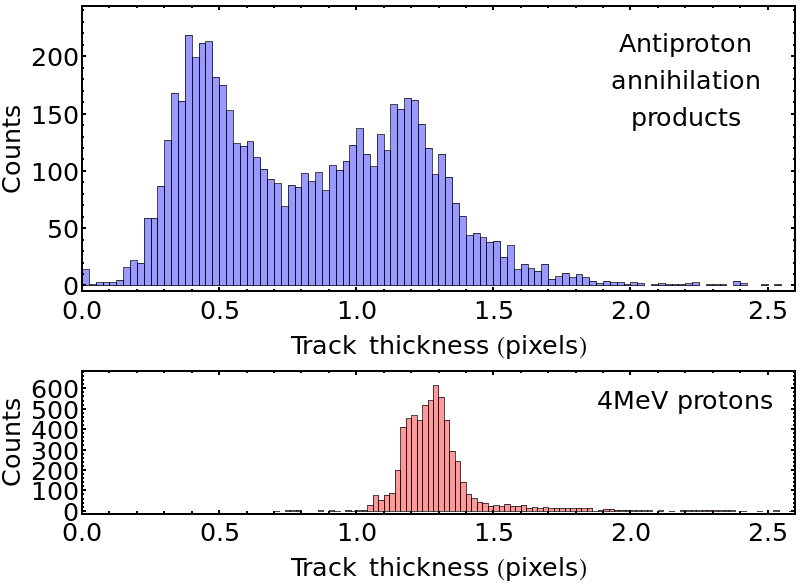}

    \caption{In blue the distribution in thickness of the \textit{tracks} in recorded antiprotons events. We attribute the higher-thickness peak of the distribution to tracks left in the sensor by protons and the lower peak to tracks left by pions. In red the distribution in thickness of tracks left by \SI{4}{\mega eV} protons within the detector, matching the topmost part of the distribution in thickness of proton tracks. We attribute the difference between the distributions mainly to the difference in energy spectrum of the protons. %Bottom: Distribution in length of the track (non ellipse-shaped) components of the annihilation vertexes. In dashed black, the expected distribution in case of perfectly isothropic emission, with pixels of ration $d = 3.5$, fitted only for tracks of length $>\SI{10}{\micro m}$
    }
    \label{fig:TrackThickness}
\end{figure*}

The vast majority of the recorded prongs are either straight lines whose thickness remains roughly constant along their entire length (\textit{tracks}) and elliptical-shaped prongs (\textit{ellipses}). We have manually flagged the number and location of every \textit{track} and \textit{ellipse} within the dataset. We have run a least-square fit algorithm to precisely determine the average thickness in pixels of all \textit{tracks} in the dataset (see Section \ref{sec:trackfitting}). The distribution in thickness of tracks is bimodal (see Figure~\ref{fig:TrackThickness}, top). By applying a threshold at $\SI{1.0}{px}$ of thickness, we can classify all prongs as either \textit{thick tracks} (2382 instances, 29.9\% of recorded prongs), \textit{thin tracks} (3929 instances, 49.2\% of recorded prongs) or \textit{ellipses} (1666 instances, 20.9\% of recorded prongs). Table \ref{tab:Composition} details the composition of recorded events.

We have experimentally characterized the response of the CMOS sensor to specific charged particles to assess the most likely associations between secondaries and the observed prongs. We have exposed the sensor to the decay products of an $^{241}$Am source positioned at a \SI{20}{\degree} angle with respect to the surface and observed that the tracks left by the \SI{5.4}{\mega eV} alpha particles appear as elliptical-shaped prongs with a consistent size of $\qtyproduct{4 x 12}{px}$. We have also exposed the sensor to a proton beam of \SI{4}{\mega eV} of energy impinging at \SI{10}{\degree} angle with respect to the surface and observed them to produce linear tracks of consistent thickness in the images. We applied the same algorithm as described in Section~\ref{sec:trackfitting} to determine the thickness of the proton tracks and found them to have thickness comprised between 1.0 and \SI{1.5}{px} (see Figure \ref{fig:TrackThickness}, bottom).

Based on these observation we attribute \textit{thick tracks} to protons, \textit{thin tracks} to charged pions and the \textit{ellipses} to deuterons, tritium, alpha particles or heavier fragments. This assignment is supported by the fact that similar features can be observed in both experimental and simulated $\pbar$ annihiliation events on the Timepix3 ~\cite{holmestad2018data}. Furthermore, these three species are known to produce similar formations within cloud chambers~\cite{balestra1985experimental}, albeit three orders of magnitude larger in size as the density inside of a cloud chamber is about a thousand times lower than that of crystalline silicon.

We have further validated this prong identification by comparing experimental and simulated data. We have tallied the observed multiplicities per vertexed annihilation event, as reported in the last row of Table~\ref{tab:Elements}. We have then employed FLUKA~\cite{FLUKA}, which includes a state-of-the-art model of antiproton annihilation, to simulate $\pbar$-Si annihiliations and count the multiplicities of secondary particles that are emitted towards the half-space containing the sensor active volume. This results in the ``FLUKA base prediction'' in Table~\ref{tab:Elements} which, however, significantly deviates from the experimental data. This is due to selection bias, since not all annihilation events can be vertexed, and to missed prong identifications, e.g.\ when a track is too short or due to occlusion from other prongs. To account for these effects, we have generated a synthetic dataset employing 1000 FLUKA-generated events. For each event a collection of tracks and ellipses is printed, according to the species of the secondary particles and their emission direction. We have manually flagged the events as traceable or untraceable and, for the latter, counted the number of tracks and ellipses that could be recognised. This results in the ``FLUKA occlusion corrected'' prediction, which shows a reasonable agreement with the observed data. Furthermore, by counting the traceable events we estimate the vertex reconstruction efficiency to be around 
% 71\%.
70\%.

%Figure \ref{TrackThickness}, bottom, shows the distribution of track length $\ell$ in antiproton annihilation events recorded during the course of our experiment. Assuming that the particle trajectories are isothropically distributed half-lines and the tracks are the intersections with the detector active region, modelled as the volume comprised between two parallel planed at distance $d$, the track length length distribution would be
% $$
%     p(\ell) ~=~ \frac{\ell}{d^2} \left( 1 + \frac{\ell^2}{d^2} \right)^{-3/2} .
% $$
%$
%    p(\ell) = \ell/d^2 ( 1 + \ell^2/d^2 )^{-3/2} .
%$
% with $d$ being the aspect ratio of the sensitive volume of the detector. 
%\davide{In this equation $d$ is not the aspect ratio of the pixels, but the depth of the sensitive region. The two definition coincide only if $\ell$ is measured in pixels.}
%An effective value $2.5 < d < 3.5$ is more appropriate in this context as energy deposited in the proximity of the sensitive volume will still contribute to the readout, as seen in~\cite{NatureCMOS,Funsten1}. As visible from the image, a portion of the shorter tracks is missing from the detection, this is due to the fact that shorter tracks can be more easily missed from counting due to them being hidden behind larger features of the annihilation. We estimate the portion of tracks missing to be between 40\% and 55\% depending on the adopted value of $d$ and the portion of the spectrum being used to fit the model.

\begin{table}
    \centering
    \begin{tabular}{|l|ccc|}
        \hline
        \textbf{Prong type} & \textbf{Ellipse} ($\boldsymbol{\alpha}$) & \textbf{Thick track} ($\boldsymbol{p}$) & \textbf{Thin track} ($\boldsymbol{\pi}$)\\ \hline
        Particle identification   & alpha/fragment & proton & charged pion\\
        FLUKA base prediction     & 1.91 & 0.85 & 1.37 \\
        FLUKA occlusion corrected & 0.75 & 0.90 & 1.46 \\ \hline
        Selection criteria        & hand tagging & width $\geq 1.0$ px & width $< 1.0$ px\\
        Experimental data         & 0.64 & 0.91 & 1.50 \\ \hline
    \end{tabular} \vspace{2mm}
    \caption{Prongs multiplicities as observed in the experimental dataset and in FLUKA-based predictions. For simulated data it is assumed that charged pions result in thin tracks, protons in thick tracks, all fragments containing two or more nucleons result in ellipses, while all other particles are ignored. The base FLUKA predictions have to be corrected for selection bias in the vertexing procedure and for occlusions among prongs, see main text for details.}
    % \davide{Uncertainties of the numerical values are not provided as these are dominated by systematics that cannot be fully modelled, but are expected to be in the order of 10\%.}
    % and their prominence in the data. Among the annihilation products predicted by the simulators, protons and charged pions were counted as sources of tracks while alpha particles, deuterons and atomic fragments were counted as ellipses. The values reported in the table already take into account the coverage of only half of the solid angle by the detector. Since not all components might be distinguishable within an antiproton annihilation event, the expected observable multiplicities within recognizable events will differ from the simple count of decay products. Predictions of the observed multiplicities were obtained by rendering events based on the simulator predictions and then classifying the resulting synthetic events by hands and are reported in the table.
    \label{tab:Elements}
\end{table}

%\note{From Helga's thesis we do not get the percentage of each species but its "multiplicity". Pion: 2.81 (fluka) 2.21 (geant), Proton 2.04 (fluka), 0.66 (geant), alpha 1.44 (fluka) 0.12 (geant), heavier stuff 1.2 (fluka) 0.65 (geant)}

\subsection{Annihilation vertex reconstruction}

\begin{figure*}[t]
    \center
    \includegraphics[width=0.8\textwidth]{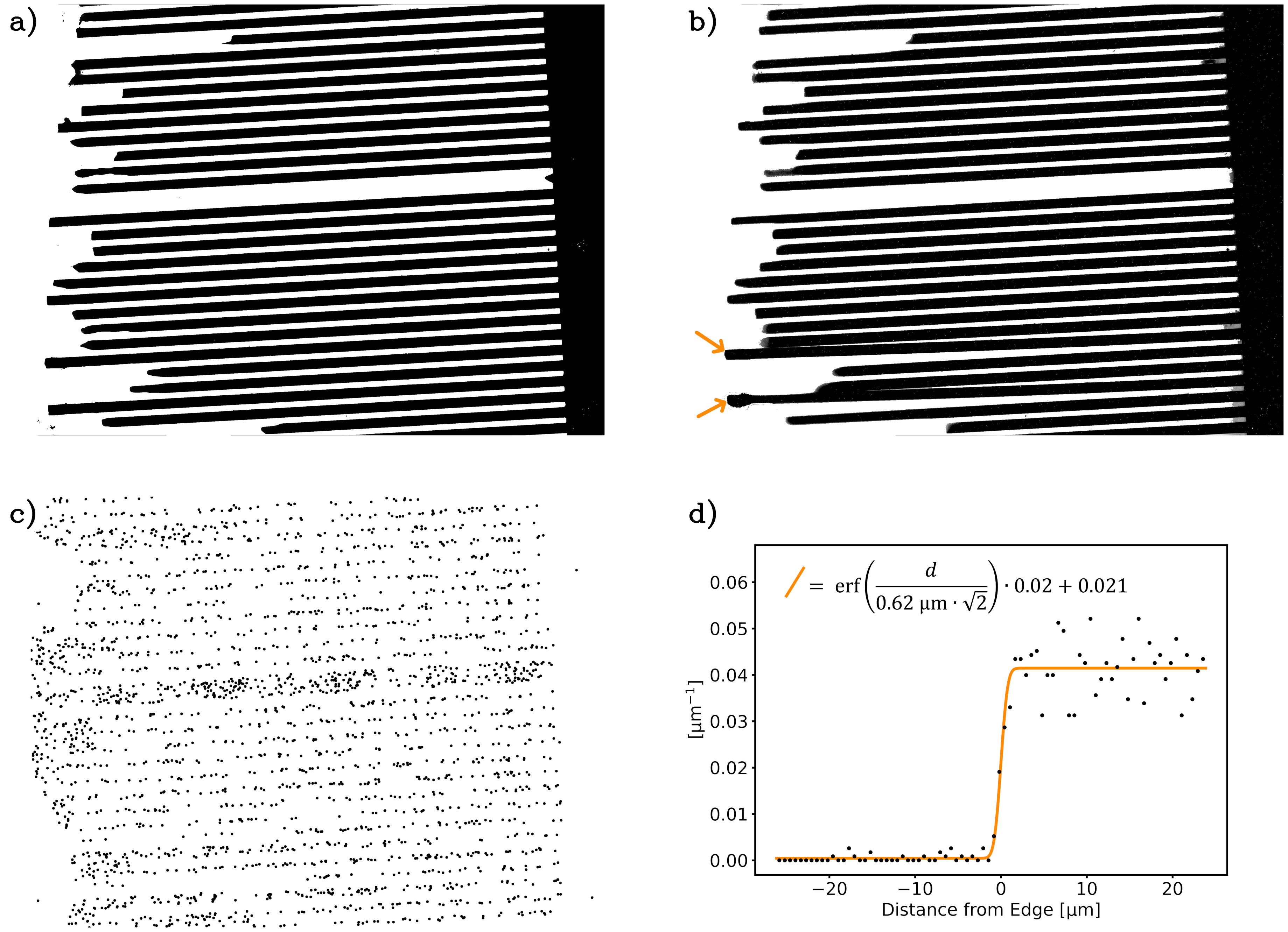}
    \caption{a) Optical image of the silicon nitride grid acquired using a point light source before installing the setup in vacuum. b) A similar image obtained by exploiting the light emitted by a vacuum gauge in the apparatus. Arrows indicate the two grid teeth that were bent by the adhesive deforming under vacuum. c) The reconstructed position of antiproton annihilations recorded with the grid installed onto the sensor. Towards the left side two distinct areas nearly devoid of annihilations events can be seen. These are the regions where the thermoplastic adhesive keeping the grid in place was applied. d) A histogram of the distance of the annihilations events from the nearest edge. Overlayed in orange is the best fitting error function found via maximum likelihood.}
    \label{fig:Grid}
\end{figure*}

We have attempted algorithmic fitting of the annihilation events, as previously done for Timepix3-based detections~\cite{holmestad2018data}. We have developed a model that predicts the intensity recorded by each pixel in the event based on a parameterization of the event (position of the vertex, number and positions of the prongs); the fitting algorithm then finds the parameterization that minimizes the $\ell^2$-distance between the recorded data and the prediction. Despite promising initial results from this approach (reaching approximately \SI{2.5}{\micro\meter} resolution), we found out that humans actually achieve higher accuracy.

We have then assigned the task of determining the position of the annihilation vertex in each of the 2601 recorded events to 8 researchers within our team. The reconstructed positions are shown in Figure~\ref{fig:Grid}c. The volunteers were provided with a graphical interface that prompted them with a $\qtyproduct{128 x 128}{px}$ portion of the image containing an annihilation event. The interface allows marking the vertex position with a precision of $1/4$ of a pixel and a tool to trace straight lines from the vertex outwards, easing the evaluation of the alignment of the vertex position with the tracks in the event. Reconstructions of a vertex position made by different volunteers for the same event differ typically by less than one pixel, with a standard deviation of $\SI{0.9}{px}$ in both the $x$ and $y$ direction. We have averaged the output of the volunteers to determine a best estimation of the vertex position. As long as no significant systematic bias is present, the accuracy of the average position should scale as $0.9/\sqrt{n}\,\SI{}{px}$, where $n$ is the number of independent vertex reconstructions. For $n=8$ this would theoretically result in a resolution of $\SI{0.32}{px}$ ($\SI{0.36}{\micro\meter}$).

\subsection{Grid positioning and vertexing accuracy}

\begin{figure*}[t]
    \center
    \includegraphics[height=100pt]{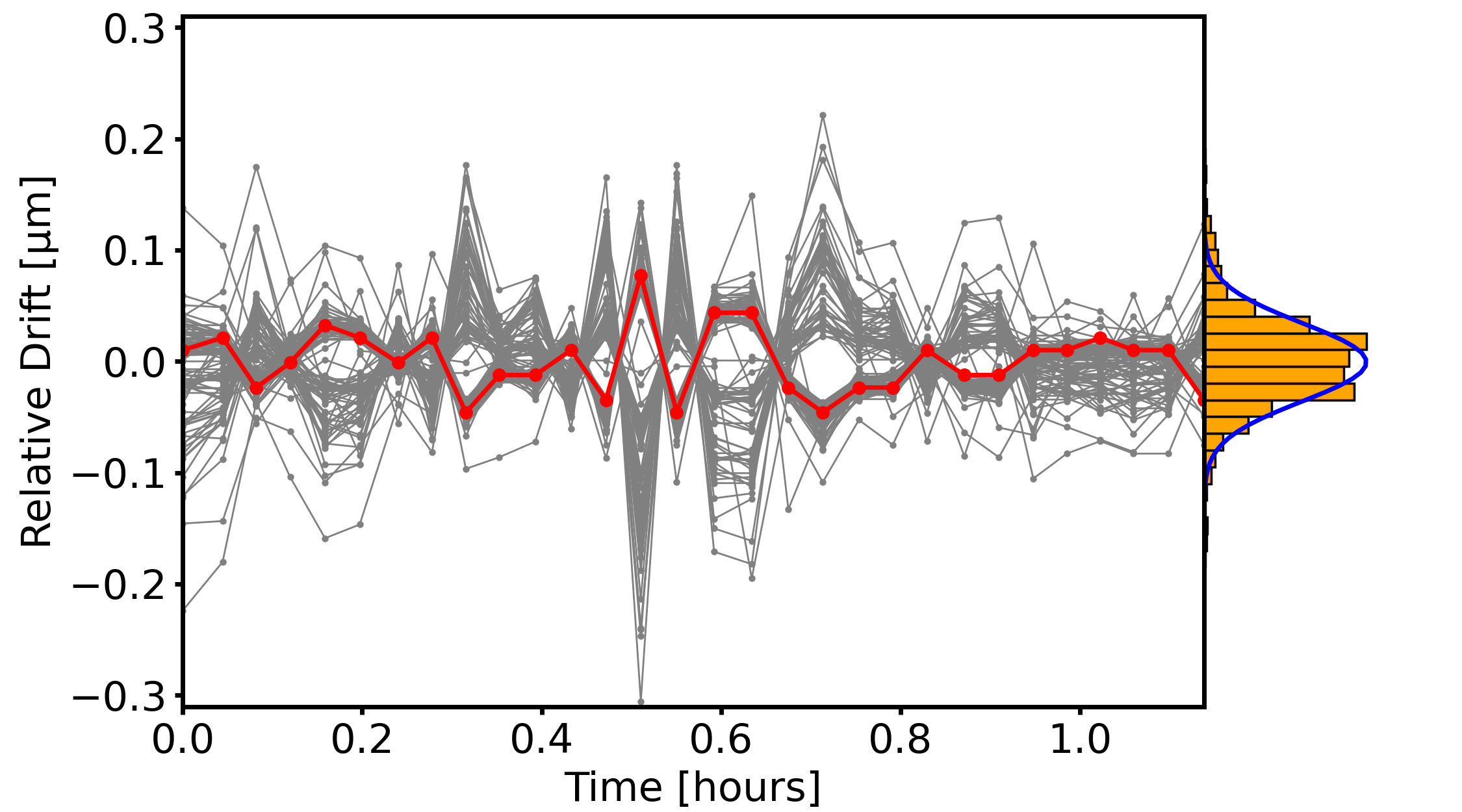}%
    \!\!\!\!\!\!\!%
    \includegraphics[height=100pt]{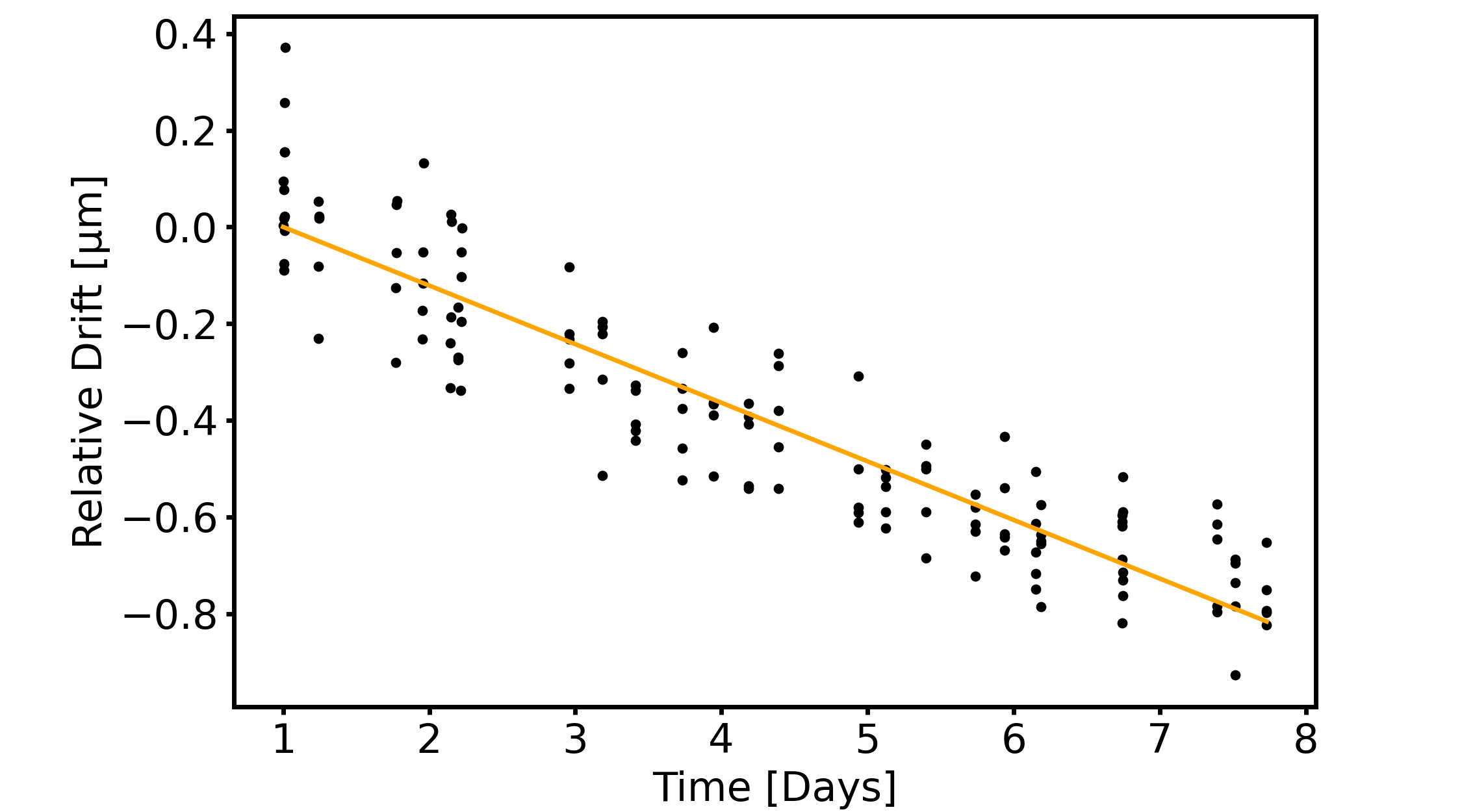}    
    \caption{Left: Relative drift of the endpoints of the 50 grid teeth as determined by the fitting algorithm described in appendix \ref{app:GridFitting} over the course of one hour. One of the edges has been highlighted in red to better show the typical progression of the algorithm output over time. The fluctuations in the so reconstructed coordinates are below \SI{0.12}{\micro\meter} RMS for all the grid vertices - with median being \SI{0.031}{\micro\meter} RMS. Right: relative position of one grid vertex tracked over the course of the 8 day measurement campaign. The linear fit used to compensate for the grid movement in the data analysis is shown in orange.}
    \label{fig:GridFittingStability}
\end{figure*}

We have performed a bias-free assessment of the crowd-based vertexing accuracy via a procedure akin to a optical transfer function measurement, a method used to determine the resolution of digital camera sensors~\cite{Reichenbach_Resolution}. Namely, we have occluded part of the sensor so that it cannot be reached by the impinging antiprotons and observed the sharpness of the transition in the density of annihilation events between occluded and non-occluded portions of the sensor. We have placed a silicon nitride grid of thickness \SI{125}{\micro\meter} consisting of teeth with a pitch of \SI{100}{\micro\meter} and openings of \SI{40}{\micro\meter} directly onto the sensor surface. In order to cover most of the active surface of the sensor while not interfering with the bonding wires, the grid was cut to size using a hardened steel scribe and then glued in place with 
thermoplastic adhesive.
% bricolage hot glue.
The grid was brought as close as possible to the sensor by observing the grid shadow as imaged by the device in real-time using visible light and employing the sharpness of its shadow to assess its proximity to the surface. An image of the grid, obtained from the sensor itself after it was installed in vacuum is shown in Figure~\ref{fig:Grid}b). In the image, two of the grid teeth are bent upwards, which was not the case before air was pumped out of the chamber. Most likely, this is due to mechanical stress caused by air pockets deforming the adhesive when vacuum was applied and had to be factored in during the data analysis.

The position of the grid relative to the sensor was measured between each run by using the residual light given off by a Pirani gauge, which was then switched off to allow the sensor to operate in darkness. We determined the position of each tooth of the grid with sub-pixel accuracy by using the algorithm described in Section~\ref{app:GridFitting}. By fitting multiple images acquired in quick succession over the course of one hour (see Figure \ref{fig:GridFittingStability}, left) we found the algorithm to be stable, with a standard deviation of the results below \SI{0.12}{\micro\meter}. Tracking of the grid position over the course of the measurement campaign reveals a linear drift in the order of \SI{0.1}{\micro\meter\,per\,day} (see Figure \ref{fig:GridFittingStability}, right). This effect is most likely due to the combination of the mechanical tension introduced in the adhesive during evacuation and the adhesive softening under the heat load of the detector during operation, which reaches temperatures in excess of \SI{60}{\celsius}.
To account for this drift, the position of the grid at any given time was interpolated using a linear regression model. This technique allowed in-situ  calibration of the grid position, ensuring consistency throughout the measurement campaign.

Antiprotons annihilating onto the grid also sometimes produce tracks, but these have a different appearance from the events shown in Figure~\ref{fig:StarShape}, allowing their exclusion of these events from the analysis. In these cases the annihilations are offset by \SI{125}{\micro\meter} from the surface of the sensor, resulting in disjoint tracks that do not touch at a vertex. For each crowd-reconstructed vertex the signed distance $d$ to the closest tooth edge was computed and the vertices having $\vert d \vert \leq \SI{25}{\micro\meter}$ were collected. The resulting distribution (in Figure~\ref{fig:Grid}d) resembles, as expected, an error-function distribution~\cite{Reichenbach_Resolution} and is modelled as $f(d)=b + n\,\text{erf}[d/ (\sqrt{2}r)]$ with fit parameters $b,n,r$, where $r$ represents a upper bound of the resolution that is acheivable with the sensor. Via a maximum likelihood fit we obtain $r = 0.62^{+0.40}_{-0.22}\,$\SI{}{\micro\meter} for crowd-based vertexing, while $r = 0.79^{+0.35}_{-0.24}\,$\SI{}{\micro\meter} is consistently obtained when the vertexing from a single volunteer is employed.

\subsection{Sensor resilience}

As the measurement was carried out, the emergence of dead pixels has been observed. Malfunctioning pixels in the detector manifest characteristically as pixels whose measurement baseline is uncharacteristically high, typically exceeding one fourth of the total dynamic range and often saturating it completely. These pixels can be easily identified in the background images taken before each antiproton implantation.
% and their emergence tracked.

The pristine sensor contained about 2500 dead pixels (one in 3200), defined as pixels whose baseline exceeded \SI{400}{lsb} (least significant bits), when acquiring in complete darkness with the same settings employed to detect antiprotons, with the maximum possible readout being \SI{1023}{lsb}.
%Quoting an exact number is meaningless as changes in the sensor temperature shift the pixels baseline, which regardless of the threshold used to discriminate dead pixels creates an apparent aging of the sensor as it heats up. This apparent aging is reversed by cooling down the detector.
%On top of that 
As we implant antiprotons in the device we observe pixels which get permanently damaged by the radiation impinging on the detector. Due to the limitations of the beam optics during this run, most of the energy deposited in the sensor during our measurements did not come from antiprotons annihilating on the detector, but instead from antiprotons annihilating elsewhere in the chamber whose annihilation products traversed the active volume of the sensor.
%To evaluate the contribution of these events to the aging of the sensor we first computed from the collected the he average signal deposited onto the sensor by an antiproton annihilating onto its surface
We evaluated their contribution to the total irradiation of the sensor by integrating the signal that they produced in the sensor and then dividing it by the average signal induced by an antiproton (\SI{38}{\kilo lsb}). Using this conversion we observed the appearance of one additional permanently damaged pixel every four equivalent antiprotons implanted into the sensor. This rate of degradation is completely negligible for its use in \aegis where an integrated flux of a few thousand antihydrogen atoms is expected during the course of the entire experiment. In fact the detector can be foreseen to be suitable to support experiments requiring an antiproton/antihydrogen flux between two and three order of magnitude larger.

\section{Discussion} \label{sec:discussion}

We have demonstrated the possibility of detecting antiproton annihilations using a commercial mobile camera sensor. We have shown that, employing crowd-sourced vertex reconstruction, a position accuracy of $0.62^{+0.40}_{-0.22}\,$\SI{}{\micro\meter} can be achieved, a 35-fold improvement over the previous state-of-the-art of $\SI{22}{\micro\meter}$ for real-time detection~\cite{holmestad2018data}. Making use of FLUKA simulations, we can estimate that around 
% 71\% 
70\%
of the annihilating antiprotons result in traceable events. Differently from previous use of the same detector for positrons~\cite{Berghold2023}, most of the energy deposited in the sensor by antiprotons comes from the rest mass of the annihilating particles. As such we expect no lower limit in the antiparticle kinetic energy in order to vertex the annihilation event and, thus, this sensor could be directly employed to detect cold antihydrogen in \aegis gravity measurement. 

Importantly, the sensor is by design sensitive to visible light, which enables the accurate determination of its position relative to the moir\'e deflectometer grids by means of optical alignment. In essence, photons can serve as test particles in a gravitational experiment: being their deflection by Earth's gravitational field negligible, they can be used to ascertain the expected arrival point of particles in the absence of gravitational acceleration\footnote{Electrons with keV-scale energy might also be employed in alternative to photons.}. Thus, by alternating between optical calibrations and antihydrogen shots a direct differential measurement of particle arrival positions at the detector can be performed. The use of a single device both for optical imaging and for antimatter detection ensures that all measurements occur within the same physical frame of reference, minimizing systematic errors. Furthermore, any drift in the position of the deflectometer grids can be detected and accounted for, as demonstrated in this work.

Following the measurements results presented here, this sensor technology has been selected as the basis for the development of a bespoke detector to be employed within the \aegis deflectometer. The detector will feature a large sensitive area (around $\qtyproduct{5.8x5.7}{\centi\meter}$), consisting of 48 individual CMOS sensors arranged in a compact rectangular tessellation, which is expected to collect 56\% of the antihydrogen atoms that pass through the deflectometer grids. The sub-micrometric accuracy and the in-situ calibration capabilities of this detector technology should allow  measuring the gravitational acceleration to $1\%$ precision, which is the \aegis scientific goal, under relaxed experimental constraints.
%For instance, if the anithydrogen atoms are produced at a temperature of $\SI{100}{\milli\kelvin}$, then 500 $\Hbar$ annihilations have to be detected in order to reach the accuracy goal~\cite{aghion2013prospects, pistillo2015emulsion}.
Ultimately, we expect that this detector technology could enable in future upgrades to significantly exceed the $1\%$ precision goal in the measurement of $g_{\Hbar}$.

As a final outlook, this sensor technology shows great promise for a broad range of applications, beyond the detection of antiprotons, due to its sensitivity to both energetic charged particles and light. We have already demonstrated that it can detect positrons~\cite{Berghold2023} and extreme-ultraviolet (XUV) light~\cite{XUVPeparation} and we expect that it may be employed in biomedical imaging~\cite{Medipix}, in visible and XUV light spectrometry, as well as in high-resolution particle tracking.

\section{Methods} \label{sec:methods}

\subsection{Hardware setup}

The apparatus used in~\cite{Berghold2023}, seen there in Figure 1, was repurposed for this experiment. The flange there depicted was installed horizontally onto a $45\degree$ offshoot of the injection line of \aegis, facing in the direction of the \SI{5}{T} magnet of the experiment. Between the detector and the injection line a \SI{10.4}{\centi m} long, \SI{4}{\centi m} wide restriction terminating in a \SI{2}{\centi m} wide aperture on the side of the detector was installed, while on the side of the injection line a pneumatically actuated gate valve was placed. The portion of tube containing the detector and the restriction were both evacuated using separate turbo pumps, reaching a vacuum of $2\cdot 10^{-7} \SI{}{\milli bar}$. The gate valve was opened only briefly during the injection of the antiprotons, to limit the contamination of the \aegis vacuum from the sensor readout system, which was not designed to reach higher levels of vacuum.

Antiprotons were captured from the Extra Low Energy Antiproton ring (ELENA) by the main \aegis apparatus and then dumped back into the injection line using a purposely implemented procedure in the experiment control system~\cite{Circus}. The injection line has been instrumented with two pairs of einzel lenses and a pair of deflecting electrodes, to allow the formation of a beam of antiprotons in two offshoot lines onto one of which the detector apparatus was installed. Unfortunately many of the electrodes were not available during the course of this experiment, so only minimal focusing of the re-emitted antiprotons was possible. This was not an issue, as the high number of antiprotons available from \aegis (estimated $4\cdot 10^6$) allowed to record on average $1.9$ events on the detector in each run, thus allowing for the execution of the experiment.

As in~\cite{Berghold2023}, images from the sensor were transmitted outside the vacuum chamber through a WiFi bridge. Syncronization of the valve actuation and minimization of its opening time was achieved acoustically by installing a microphone onto the valve and matching the recorded audio to the master clock of the experiment.

\subsection{Track fitting algorithms}
\label{sec:trackfitting}

The algorithm processes a $128 \times 128$ pixel section of the image that includes a preselected annihilation event and outputs the widths of all present tracks. Since tracks and ellipses in each annihilation event often overlap, it is essential to estimate their widths jointly to minimize biases. As auxiliary input, the number of tracks and ellipses that are present in the image has to be provided, as well as a preliminary estimation of positions of these tracks and ellipses. Each track position is identified by its two endpoints, $\vec{p}_0=(x_0,y_0)$ and $\vec{p}_1=(x_1,y_1)$, while each ellipse is identified by its center $\vec{p_c}=(x_c,y_c)$ and by two distinct points on its boundary, $\vec{p}_a=(x_a,y_a)$ and $\vec{p}_b=(x_b,y_b)$, so that the boundary can be parameterized as $\vec{p}(\phi) = (\vec{p}_a-\vec{p}_c) \sin (\phi) + (\vec{p}_b-\vec{p}_c) \cos (\phi)$, for $\phi \in [0,2\pi)$. The initial positions are a human inputs and must be accurate within 2 pixels.

The widths of the tracks are estimated by modeling the expected appearance of the annihilation event based on a few underlying parameters and then identifying the parameters that best fit the observed data. Specifically, an annihilation event containing $n_\text{ell}$ ellipses and $n_\text{tr}$ tracks is described via the parameter set $\mathcal{T}=\{(\vec{p}_{a,i}, \vec{p}_{b,i}, \vec{p}_{c,i})\}_{i=1}^{n_\text{ell}} \bigcup \{(\vec{p}_{0,i}, \vec{p}_{1,i}, w_i)\}_{i=1}^{n_\text{tr}}$, where $w_i>0$ is the width of the $i$-track.

Given the parameterization $\mathcal{T}$ of an event, a model of the intensities of the pixels is obtained as follows. The intensity $I(\vec{r})$ of the pixel with coordinates $\vec{r}=(x,y)$, assumed to be at the center of the pixel, ranges from 0 (black) to 1 (white). The distance $d$ of each pixel to the boundary of each ellipse is numerically computed, with $d>0$ ($d<0$) indicating points inside (outside) an ellipse. The pixel intensities due the $i$-th ellipse are 
\begin{align}
    I_i^{\text{ell}}(\vec{r}) =
    \begin{cases}
    1   & \text{if}~d < 0  \\
    1-d & \text{if}~0 \leq d < 1\\
    0   & \text{if}~d \geq 1 .
    \end{cases}
\end{align}
Similarly, the distance $d$ of each pixel to each track is also numerically computed (being a track identified by a segment, this distance cannot be negative). The pixel intensities due the $i$-th track are
\begin{align}
    I_i^{\text{tr}}(\vec{r}) =
    \begin{cases}
    1   & \text{if}~d < w_i\\
    1-d & \text{if}~w_i \leq d < w_i+1\\
    0   & \text{if}~d \geq w_i+1 .
    \end{cases}
\end{align}
This differs from anti-aliasing algorithms for drawing lines in computer graphics, but it more closely resembles the detector physics and provides a better fit to the tracks. Finally, the intensity of each pixel in the model is obtained by summing the individual contributions and clipping the result to one, i.e.\  $I_\mathcal{T}(\vec{r}) = \min\left(1, \sum_{i=1}^{n_\text{ell}} I_i^{\text{ell}}(\vec{r}) + \sum_{i=1}^{n_\text{tr}} I_i^{\text{tr}}(\vec{r}) \right)$.

The $\ell^2$ distance between the modelled intensities $I_\mathcal{T}$ and the observed intensities $I_\text{data}$ is computed as 
\begin{align}
    \ell^2(\mathcal{T}) = 
    \sum_{\vec{r}\in\mathcal{R}} \big\vert I_\mathcal{T}(\vec{r})-I_\text{data}(\vec{r})\big\vert^2  ,    
\end{align}
where the dead pixels are excluded from the summation domain $\mathcal{R}$. The parameters in $\mathcal{T}$ are then varied within $2$ pixels from their preset values and a local minimum of $\ell^2(\mathcal{T})$ is obtained via a Nelder-Mead optimiser. From our analyses, we expect these local minima to be also global ones. The values of the track thicknesses are extracted from the value of $\mathcal{T}$ found by the optimiser.

\subsection{Grid fitting algorithms}
\label{app:GridFitting}

Given as input an image taken by the sensor and set of segments providing a preliminary estimation of the positions of the edges of the teeth, the algorithm outputs a refined estimation of the position of these edges. The preliminary position estimation is a human input.

Let $k$ be the number of tooth edges (typically, twice the number of teeth) and for each $i \in \{1,\ldots,k\}$ let the points $(x_{0,i},y_{0,i}), (x_{1,i},y_{1,i})$ be the endpoints of a segment indicating approximately the position of the $i$-th edge. It is assumed that the segment is almost horizontal (i.e., $\vert x_{1,i} - x_{0,i} \vert \gg \vert y_{1,i} - y_{0,i} \vert$) and, without loss of generality, that $x_{1,i} > x_{0,i}$. The equation of the $i$-th segment is
\begin{align}
    y_i(x) =
    \frac{y_{1,i} - y_{0,i}}{x_{1,i} - x_{0,i}} x + 
    \frac{x_{1,i} y_{0,i} - x_{0,i} y_{1,i}}{x_{1,i} - x_{0,i}}
    \qquad \text{for} ~ x_{0,i} \leq x \leq x_{1,i} .
\end{align}
A band of pixels having width $2w$ around each segment is considered, given by the pixels having coordinates $(x,y)$ within vertical distance $w$ from the $i$-th segment. Let their intensities in the image be $I_i(x,y)$. We define $\bar{I}_i(x, y)$ as a Gaussian smoothing of these intensities along the vertical direction within a band, i.e.
\begin{align}
    \bar{I}_i(x, y) =
    \frac{\sum_{y'\in {B_i}(x)} I(x,y')\, e^{\frac{(y-y')^2}{s^2}}}
         {\sum_{y'\in {B_i}(x)}           e^{\frac{(y-y')^2}{s^2}}} 
    \qquad B_i(x) = \{ y: ~\vert y-y_i(x)\vert \leq w\}.
\end{align}
Let $p$ be a real indicating the portion of image covered by the grid and $N(I)$ the number of pixels having intensity smaller or equal than $I$, with $N(\infty)$ denoting the total number of pixels in the image. First, a threshold value $I_p$ is determined as the largest intensity such that $N(I_p) < p \cdot N(\infty)$. Then the equation
\begin{align}
    \bar{I}_i(x_i, y) = I_p
    \label{eq:threshold}
\end{align}
is solved numerically in $y$ for all $i\in \{1,\ldots,k\}$ and for all $x_i \in \{x_{0,i}, \ldots, x_{1,i}\}$. If the solution to Eq.\eqref{eq:threshold} is not unique, the point is discarded and a warning is raised. For each edge $i$, a collection of points $\mathcal{C}_i = \{(x_i,y_i): \bar{I}_i(x_i, y_i) = I_p \text{~and~} y_i \text{~is~unique} \}$ is created. Finally, for each edge $i$, the output is a straight line obtained via linear regression of $y$ as a function of $x$ over the collection $\mathcal{C}_i$.

In our case, the tunable parameters of the algorithm were set to $p = 0.6$, $w=10$ and $s^2 = 3$.

\section{Acknowledgements}

We would like to thank Michael Fu\ss{}eder and Thomas Schwarz-Selinger from  the tandem accelerator lab of the Max Planck Institute for Plasma Physics (IPP) for granting us the beam time necessary to characterize proton interactions with the sensor. Also funding from 
%
% Giovanni Consolati / Roberto S. Brusa / Ruggero Caravita / Nicola Zurlo
Istituto Nazionale di Fisica Nucleare; 
%
% Michael Doser 
the CERN Fellowship programme and the CERN Doctoral student programme; 
%
% Carsten Welsch
the EPSRC of UK under grant number EP/X014851/1;
%
% Antoine Camper
Research Council of Norway under Grant Agreement No. 303337 and NorCC;
%
% Natali Gusakova
CERN-NTNU doctoral program;
%
% Georgy Kornakov
the Research University – Excellence Initiative of Warsaw University of Technology via the strategic funds of the Priority Research Centre of High Energy Physics and Experimental Techniques;
the IDUB POSTDOC programme;
the Polish National Science Centre under agreements no. 2022/45/B/ST2/02029, and no. 2022/46/E/ST2/00255, and by the Polish Ministry of Education and Science under agreement no. 2022/WK/06;
%
% Tomasz Sowinski
% the Polish Ministry of Education and Science on the basis of agreement no. 2022/WK/06; N.B.: same as previous line
%
% Name
Marie Sklodowska-Curie Innovative Training Network Fellowship of the European Commission's Horizon 2020 Programme (No. 721559 AVA); 
%
% Name
European Union's Horizon 2020 research and innovation programme under the Marie Sklodowska-Curie grant agreement ANGRAM No. 748826; 
%
% Saiva Huck / Lisa Gloggler / Tim Wolz
Wolfgang Gentner Programme of the German Federal Ministry of Education and Research (grant no. 13E18CHA); 
%
% Name
European Research Council under the European Unions Seventh Framework Program FP7/2007-2013 (Grants Nos. 291242 and 277762); 
%
% V. Petracek Czech Technical University in Prague
the European Social Fund within the framework of realizing the project, in support of intersectoral mobility and quality enhancement of research teams at Czech Technical University in Prague (Grant No. CZ.1.07/2.3.00/30.0034);

\bibliography{bibliography}

\end{document}